# A Compact and Discriminative Feature Based on Auditory Summary Statistics for Acoustic Scene Classification

*Hongwei Song[1], Jiqing Han[1], Shiwen Deng[2]*

[1] Harbin Institute of Technology, China
[2] Harbin Normal University, China

15B903035@hit.edu.cn, jqhan@hit.edu.cn, dengswen@gmail.com

## Abstract

One of the biggest challenges of acoustic scene classification (ASC) is to find proper features to better represent and characterize environmental sounds. Environmental sounds generally involve more sound sources while exhibiting less structure in temporal spectral representations. However, the background of an acoustic scene exhibits temporal homogeneity in acoustic properties, suggesting it could be characterized by distribution statistics rather than temporal details. In this work, we investigated using auditory summary statistics as the feature for ASC tasks. The inspiration comes from a recent neuroscience study, which shows the human auditory system tends to perceive sound textures through time-averaged statistics. Based on these statistics, we further proposed to use linear discriminant analysis to eliminate redundancies among these statistics while keeping the discriminative information, providing an extreme compact representation for acoustic scenes. Experimental results show the outstanding performance of the proposed feature over the conventional handcrafted features.

## 1. Introduction

Acoustic scene classification (ASC), usually defined as the task of identifying the acoustic environment from the sounds they produce, has drawn attention of machine listening community recently [1]. It is one of the critical techniques that would enable machines/devices the ability of environment-awareness and there has been some real-life applications, such as context-aware services [2] and robotic navigation [3].

One of the biggest challenges of ASC is to find proper features to better represent and characterize environmental sounds. Early works have heavily borrowed features from speech and music processing fields. For instance, features used to dominate speech processing community, such as Mel frequency cepstral coefficients (MFCC) [4], Linear predictive coefficients (LPC) [5] and some low-level temporal or spectral features were widely used by ASC systems. Another example is the Const-Q Transform (CQT) [6]. It was initially designed for describing harmonic sounds like music tones, but now has been widely used as the feature for ASC tasks.

However, the acoustic properties of daily-life environments are quite different from those of speech and music signals. On one hand, environmental sounds exhibit less structure in the temporal-spectral representation. On the other hand, sounds involved in daily-life environments are more diverse and different ways of imposition of these sounds could make this problem harder. Thus, more task-adapted features should be designed for ASC tasks. Towards this goal, Rakotomamonjy et al. [7] proposed to use histogram of oriented gradients (HOG) to encode the local direction of variation on top of CQT. Bisot et al. [8] proposed to use the Subband Power Distribution (SPD) as a feature for capturing occurrences of sound events inside a scene. Beyond these feature engineering methods, some works investigated supervised or unsupervised feature learning-based methods. For instance, a recent work by Bisot et al. [9] investigated various matrix factorization techniques to learn features from CQT spectrogram. Hyder et al. [10] proposed a CNN-SuperVector system to combine the feature learning strength of CNN with the super-vector backend.

The strategy shared by the methods mentioned above is that they tried to model acoustic scenes as a whole based on temporal-spectral representations, either by handcrafted feature engineering or by feature learning. Another strategy is to treat the background and foreground of the acoustic scene separately while focusing on background modeling. For instance, the minimum statistics were extracted to model background in a scene [11] and Local Binary Patterns (LBP) with an enhanced zoning mechanism was used as a background sound texture descriptor [12]. In this work, we followed the second strategy based on the following observation: the background sounds are usually produced by a collection of co-occurrence sound events, producing a texture-like sound, which could be characterized by its temporal homogeneity.

Our work is greatly inspired by two recent works on neuroscience study of sound texture perception. The evidence from [13] suggests humans perceive texture sounds by summarizing the temporal details of sounds using time-averaged statistics. An earlier work [14] by the same group found that sound textures synthesized with the same statistics sound similar. These findings suggest these statistics underlie the distinctions among various sound textures, thus could be used for recognizing the background of acoustic scenes. Relevant to our work is [15], in which similar texture statistics were used as features for a coarse-grained, general-purpose audio track classification task, which achieved very similar accuracies to MFCC features. However, we believe these statistical features are more suitable for characterizing fine-grained sound textures, such as acoustic scene backgrounds. Experimental results support our view, with mean accuracy outperforming MFCC by a large margin on ASC datasets. Besides, we proposed to use LDA to extract an extreme compact feature with great discrimination for acoustic scene classification.

In this work, we investigated modeling the backgrounds of the acoustic scene using the auditory summary statistics. These statistics were measured upon three intermediated signals from an auditory filtering and processing module, which simulates the responses of the human auditory systems. Then a compact and discriminative feature was extracted using LDA. For experiments, we analyzed how audio segment length impact the effectiveness of the feature. And we visualized the feature extracted to show its discrimination. At last, the feature was combined with a support vector machine (SVM) classifier and tested on LITIS Rouen [7] dataset and DCASE2016 dataset [16].

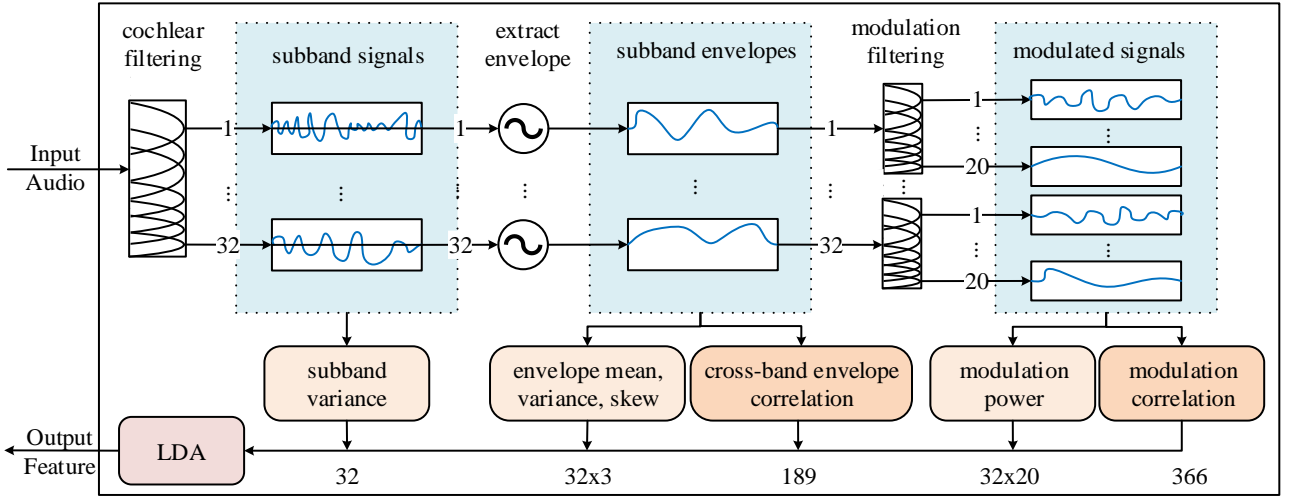

Figure 1: *Block diagram for feature extraction.*

## 2. Feature Extraction

The feature extraction process is shown in Figure 1. First, an input audio was filtered and processed to generate intermediate signals simulating responses of the human auditory system. Basically we follow the process flow of the auditory model in [14]. The intermediate signals include subband signals, subband envelopes and modulated signals, which were shown inside the dashed blocks. Then statistical moments such as mean, variance, skew and correlations were measured on these signals. Finally, these summary statistics were concatenated into a large vector and passed into a LDA module to reduce dimension. We need to point out that these statistical features were extracted from audio segments rather than the full-length audio. The main concerns and related experiments were presented in the next section.

### 2.1. Auditory filtering and processing

As shown in the upper part of Figure 1, an input audio was first passed to a 32-channel cochlear filter bank, which spans 20-10000 Hz in a log frequency scale. Thus, an input sound pressure wave was decomposed into 32 subband signals, simulating the frequency selectivity of cochlea.

Then the envelope of each subband was computed by first applying Hilbert transformation to the subband signal and magnitude of the analytic signal was token. Next, subband envelopes were raised to a power of 0.3 to simulate basilar membrane compression. The output of this step were 32 subband envelopes. The envelope signals were further down-sampled into 400 Hz for reducing computational complexity.

To replicate the modulation tuning effect in midbrain and thalamic neurons [17], a bank of 20 filters spanning 0.5–200 Hz was applied to each subband envelope, generating 20 modulated signals for each subband envelope. Therefore, there were $32 \times 20$ modulated signals in total.

### 2.2. Statistics measurement

After auditory filtering and processing, we measured the statistical moments and pairwise correlations of the intermediate signals. The rationale under statistical moments is the following: acoustic backgrounds could be characterized by its temporal homogeneity, thus could be described as distribution rather than temporal details. Hence, statistical moments (e.g. mean, variance, skew) were calculated to reflect the difference among these distributions. The correlation statistics, in contrast, each reflects distinct aspects of correlations between envelopes of different channels, or between their modulation bands. For example, some sounds are broadband (multichannel responses from cochlear responses are high) while some other are independent across channels. These dependencies or correlations were reflected in correlation statistics.

However, experimental results in [14] showed that not all the statistics were perceptually important and also there were information redundancies among these statistics. We believe perceptually important statistics would help classification. Thus, in practice, we use a subset of statistics that has proven to be perceptually important in [14]. Hence, for marginal momnets, we selected subband variance, envelope mean/variance/skew, modulation power for each frequency channel and modulation channel. For envelope correlations, we selected the six diagonal statistics from the full $32 \times 32$ envelope correlation matrix with diagonal index belonging to $\{2, 3, 4, 6, 9, 12, 17, 22\}$, resulting in 189 envelope correlation features. For modulation correlations, correlations were computed between two bands centered on the same modulation frequency but different acoustic frequencies (referred as C1 in [14]). We only calculated correlations between two nearest frequency bands and the modulation band with index $\{2, 3, 4, 5, 6, 7\}$, resulting in 366 modulation correlation features. These statistics were sufficient to reproduce the qualitative form of the full correlation matrix through correlation propagation and has proven to be perceptually sufficient [14]. Finally, these auditory summary statistics (ASS) were concatenated into a large vector (with a dimension of 1322), named ASS-vector.

## 2.3. Linear discriminant analysis

The ASS-vectors are unsuitable to be directly used as the feature in classification tasks due to its high dimensional. Hence, a feature reduction is needed. For the ASC task, the audio samples of the same acoustic scene may be recorded each in different locations. For example, recordings from different restaurants share the same label as 'restaurant'. Thus during feature reduction, we need to eliminate the variances coming from locations while keeping the variances that help distinguishing scenes. Hence, linear discriminant analysis was performed to transform the ASS-vectors into a low dimensional feature space where the between-scene covariance is maximized while within–scene covariance is minimized. More formally, the procedure is as follows: First, the between-scene and within-scene scatter matrices $S_b$ and $S_w$ were computed as:

$$S_b = \sum_{c=1}^{C} N_c (\mu_c - \mu)(\mu_c - \mu)^T \quad (1)$$

$$S_w = \sum_{c=1}^{C} \sum_{j=1}^{N_c} (w_{c,j} - \mu_c)(w_{c,j} - \mu_c)^T \quad (2)$$

where, the number of scenes (classes) is $C$, the total number of ASS-vector is $N$ and $N_c$ is the number of ASS-vectors belonging to the $c^{th}$ scene. $w_{c,j}$ is the $j^{th}$ ASS-vector from the $c^{th}$ scene. $\mu_c$ is the mean of ASS-vectors belonging to scene $c$. $\mu$ is the global mean of all the $N$ ASS-vectors in the development set.

Then our objective is to find a projection matrix $A$ so that in the projected space the following criteria is maximized:

$$J = Tr(\tilde{S}_w^{-1} \tilde{S}_b) = Tr\{(A^T S_w A)^{-1}(A^T S_b A)\} \quad (3)$$

Where $\tilde{S}_b$ is the between-scene scatter matrix and $\tilde{S}_w$ is the within-scene scatter matrix in the projected space. The values of $A$ that optimize criteria $J$ is given by eigenvectors corresponding to the largest eigenvalues of $S_w^{-1} S_b$ [18]. After getting transformation matrix $A$, for any ASS-vector $w$, we could get a low dimensional feature $v$ as

$$v = A^T w \quad (4)$$

We name this feature $v$ as ASS-LDA feature. In practice, the projection matrix $A$ was trained on training data and retained for testing. Based on preliminary experiments, the dimension of ASS-LDA feature for LITIS Rouen dataset was set to 18 while for DCASE2016 dataset, the dimension of ASS-LDA feature was set to 14.

## 3. Experimental Evaluation

### 3.1. Datasets

The experiments were carried out on two ASC datasets: the LITIS Rouen dataset [7] and DCASE2016 dataset [16]. LITIS Rouen dataset is the largest ASC dataset and since the data is unbalanced, the *mean Average Precision (mAP)* [7] over the 20 folds training-testing splits were reported, as suggested by the creators of the LITIS Rouen dataset. However, for comparison with other methods, the accuracy was also reported, which defined as the number of correctly classified samples divided by the total number of samples. DCASE2016 dataset consists of two subsets: development set (1170 files) and evaluation set (390 files). For this dataset, the mean accuracy over 4-fold cross validation of the development set and accuracy of the challenge evaluation dataset were reported. And the data from both channels were utilized in both training and testing phase for DCASE2016 dataset.

### 3.2. Segmentation and classification

#### 3.2.1. Slicing and feature extraction

An input audio was first sliced into segments before the ASS-LDA feature were extracted for each segment. The main reason why we did not extract features for the whole audio recording is based on the following two considerations: First, the effectiveness of the statistical features is based on the temporal homogeneity of scene background, which may not hold true for a full-length (30s) audio example of ASC datasets. However, an audio segment of moderate length could be thought as temporal homogeneous and thus could be modeled as a whole. Second, some rare short-term sound event may influence the statistics of a scene. For example, a high-tune female laugh on a train may have a great impact on the high-frequency statistics of the scene. But it is rather a rare event for a typical scene on a train. Thus by segmentation we could limit the influence of such rare events within one or few segments, without polluting the whole audio file.

#### 3.2.2. The SVM Classifier

The ASS-LDA feature vectors for each segment was scaled to [-1,1] before passing to a RBF Kernel, one-versus-one SVM classifier [19]. Both the training and testing phases of the SVM were all segment-based, but the final decision for each testing file was based on a majority vote mechanism.

To set parameters, for each dataset, we performed a coarse grid search by further split training set of the first fold into 80%~20% train/validation splits. Parameters giving the highest cross validation accuracy on first fold training set were retained. The final parameters for each dataset is listed in Table 1.

Table 1: *parameters for the SVM classifier.*

| Datasets | C | Gamma |
|---|---|---|
| LITIS Rouen | 4 | 2 |
| DCASE 2016 | 2 | 4 |

### 3.3. Experimental results

#### 3.3.1. The influence of segment length

The first experiment explored how segment length impact system performance on LITIS Rouen dataset. We sliced each full-length (30 seconds) audio file into segments of *seg_len* seconds, with *hop_size* set to half of the segment length, resulting in *seg_num* segments. The final decision for a full-length audio was based on majority-vote over *seg_num* segments. And the mean average precision over 20-folds cross validation setups were reported as the evaluation metric.

As illustrated in Table 2, the difference of mAP among 2s~6s were not significant and the last configuration with segment length set to 2 seconds yielded the best mAP. As expected, segment length set to 30 seconds failed to extract effective features as the temporal homogeneity premise may be broken for too long audios and the features may be polluted by some rare sound events. Thus in later experiments, segment length was set to 2 seconds for both datasets.

Table 2: *performance based on various segment lengths on LITIS Rouen dataset.*

| seg_len | hop_size | seg_num | mAP |
|---|---|---|---|
| 30s | -- | 1 | 20.2% |
| 10 s | 5 s | 5 | 92.6% |
| 6 s | 3 s | 9 | 94.1% |
| 5s | 2.5s | 11 | 94.7% |
| 3 s | 1.5 s | 19 | 94.9% |
| 2s | 1s | 29 | **95.4%** |

### 3.3.2. Visualizing ASS-vectors and ASS-LDA feature

For the second experiment, we visualized how the feature obtained carry discriminative information. The experiment was performed on the training data of the first fold of DCASE 2016. Both ASS-LDA feature and the ASS-vector were projected to 2D-dimensional space using Barnes-Hut tSNE [20]. As shown in Figure 2, The ASS-LDA were better clustered suggesting during LDA discriminative information were kept.

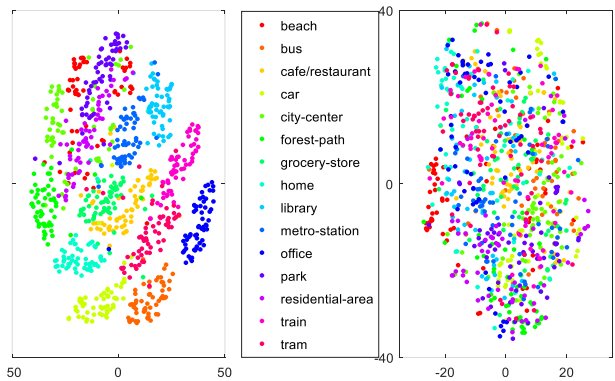

Figure 2: *Feature visualization of ASS-LDA (left) and ASS-vectors (right) using tSNE*

### 3.3.3. Results on LITIS Rouen

Table 3 summarizes some of the best-performing results on LITIS Rouen based on various features. For easy comparison, we split them into two categories: the handcrafted feature-based methods and the feature learning-based methods. As shown, although our performance was outperformed by the feature learning-based methods, we achieved significant improvements over the feature-based method without feature fusion and we got comparable results to the best-performing handcrafted feature-based method. Moreover, we provided an extreme compact feature (with a dimension of 18) comparing to other methods. This feature could be easily combined with other complementary features to improve system performance in the future.

Table 3: *Comparison with state-of-the-art methods on LITIS Rouen dataset.*

| Features | dimension | mAP | Accuracy |
|---|---|---|---|
| **Handcrafted feature-based method:** | | | |
| HOG [7] | 1536 | 92.0% | -- |
| LBP$_{8,1}$ [12] | 880 | 91.5% | 91.5% |
| HOG+SPD [8] | >1000 | 93.3% | 93.4% |
| LBP+HOG [12] | >3000 | 95.1% | 95.1% |
| ASS-LDA (ours)* | **18** | **95.4%** | **95.0%** |
| **Feature learning-based method:** | | | |
| CQT+TDL [9] | 512 | -- | 96.4% |
| DNN+LR [9] | 100 | -- | 97.1% |

### 3.3.4. Results on DCASE 2016

Since this dataset was initially released as a challenge task, many challengers tend to take a multi-model fusion strategy to push the final score higher. However, for a fair comparison, we limited our comparison among systems without utilizing multi-model fusion. For multi-model systems, the performance of the single best model was reported. Under this condition, our method achieved the state-of-the-art results on both 4-fold cross validation sets and the challenge evaluation set.

Table 4 summarizes some of the best-performing models on DCASE2016 dataset. All results reported here were directly extracted from the referenced papers. We included baseline model (MFCC + GMM [16]) from the initial challenge as our baseline. First, we included the ranked-first method of the challenge, which was an i-vector model based on boosted Multi-channel MFCC features [21]. Then ranked second method of the challenge was also included, which was a task-driven dictionary learning (TDL) model based on a CQT spectrogram. The best-performing updated version of it was reported [9]. Then a model inspired by speaker recognition was reported, in which CNN was used as a feature learner to extract feature from a SIF spectrogram [10]. At last, we include a recent work [22] investigating various deep neural network models on various features. The best-performing model of it [22] was a DNN model trained on a collection of features named Smile6k, which includes MFCC, Fourier transforms, zero crossing rate, energy, and pitch etc. as well as first and second order features.

As shown, our model significantly improved the baseline reaching an 89.5% accuracy on evaluation set. And we got superior results over various state-of-the-art models based on a variety of features, demonstrating the advantages of using ASS-LDA as representation for acoustic scenes classification.

Table 4: *Accuracy comparison with state-of-the-art methods on DCASE2016 dataset.*

| Feature | Classifier | 4CV | Eval |
|---|---|---|---|
| MFCC [16] | GMM baseline | 77.2% | 72.5% |
| MiMFCC [21] | i-vectors (CMB) | 83.9% | 88.7% |
| CQT + TDL [9] | LR | 83.8% | -- |
| SIF-CNN-SV [10] | PLDA | 81.8% | 87.2% |
| Smile6k [22] | DNN | 84.2% | 84.1% |
| ASS-LDA (ours)* | SVM | **84.6%** | **89.5%** |

## 4. Conclusion

In this paper, we proposed an LDA-based method to extract a discriminative feature from auditory summary statistics for characterizing the background of acoustic scenes. The auditory summary statistics provide a statistical representation of acoustic scenes by disregarding the temporal details of the signal. Based on these statistics, we extracted a feature that was extremely compact comparing to conventional features. With this compact feature, we achieved superior results than the best-performing handcrafted feature methods without fusion on LITIS Rouen dataset and the state-of-the-art performance on DCASE2016 dataset.

## 5. Acknowledgements

This research is partly supported by the National Natural Science Foundation of China under grant No. 61471145 and U1736210.